\documentclass[aps,prc,reprint,groupedaddress]{revtex4-2}

\usepackage{graphicx}
\usepackage{dcolumn}
\usepackage{bm}

\begin{document}


\title{Total kinetic energy release in the fast  neutron-induced fission of \ensuremath{^{237}}Np}



\author{A. Pica}

\author{A.T. Chemey}

\author{L. Yao}

\author{W. Loveland}
\affiliation{Department of Chemistry\unskip, Oregon State University\unskip, OR\unskip, 97331\unskip, Corvallis\unskip, USA}

\author{H.Y. Lee}

\author{S.A. Kuvin}
\affiliation{Physics Division\unskip, Los Alamos National Laboratory\unskip, NM\unskip, 87545\unskip, Los Alamos\unskip, USA}


\begin{abstract}
\textbf{~~~}The total kinetic energy (TKE) in the fast neutron induced fission of \ensuremath{^{237}}Np was measured for neutron energies from $_n$ = 2.6 - 100 MeV at the LANSCE-WNR facility. The post TKE release decreases non-linearly with increasing incident neutron energy and can be represented as $TKE(MeV)\;=\;(174.38\;\pm\;0.72)\;-(5.11\;\pm\;0.5821)\;\log_{10}E_n $ for E\ensuremath{_{n}} {\textgreater} 1 MeV. Analysis of the fragment mass distributions indicates that the decrease in TKE with increasing E\ensuremath{_{n}}\ensuremath{_{}}is a consequence of two factors; shell effects fade out at high excitation energies, resulting in the increasing occurrence of symmetric fission, and TKE$_{asym}$ decreases rapidly at high E$_n$.
\end{abstract}\def\keywordstitle{Keywords}

\keywords{Total kinetic energy, nuclear fission, mass yield, Np-237}


\maketitle

\section{INTRODUCTION}
The fission process emits an enormous amount of energy per event, of the order of 200 MeV per nucleus for actinides. The majority of the energy released in the fission of actinide nuclei appears in the form of the kinetic energy of the two fission fragments. According to the conservation of momentum, the partitioning of energy between the two fragments is dependent on their masses; the lighter fragment garners more kinetic energy than its heavier counterpart. The sum of the two fragments’ kinetic energy is the total kinetic energy (TKE). The TKE is an important fundamental feature of fission, and studies of TKE can help to elucidate some of the physics behind the large scale collective motion of the nucleus. \newline
\indent	Two dynamics dictate the magnitude of the TKE release. The first is the large Coulomb repulsion between the proto-fragments at the scission point,  and the second is the conversion of the collective energy of the nucleus to kinetic energy of the fragments as the nucleus transitions from the saddle point to the scission point. Based on these features, we can write TKE as a function of the fragments charge and radius, assuming R = $r_o\;^{1/3}$.

\let\saveeqnno\theequation
\let\savefrac\frac
\def\dispfrac{\displaystyle\savefrac}
\begin{eqnarray}
\let\frac\dispfrac
\gdef\theequation{1}
\let\theHequation\theequation
\label{dfg-b55e055b7286}
\begin{array}{@{}l}\begin{array}{l}TKE\;=\frac{Z_1Z_2e^{2}}{r_0(A_1^{1/3}+A_2^{1/3})}\\\end{array}\end{array}
\end{eqnarray}
\global\let\theequation\saveeqnno
\addtocounter{equation}{-1}\ignorespaces 
Equation 1 is derived from the Coulomb forces acting on the fragments. While this equation is reasonable at making first order approximations of TKE, it slightly overestimates the impact of Coulomb repulsion because the equation assumes the proto-fragments are perfect spheres at the scission point. In reality, they are highly deformed. Such deformation means the proto-fragments are more distant from each other at the scission point, and therefore experience less Coulomb repulsion. A more widely used equation is
\let\saveeqnno\theequation
\let\savefrac\frac
\def\dispfrac{\displaystyle\savefrac}
\begin{eqnarray}
\let\frac\dispfrac
\gdef\theequation{2}
\let\theHequation\theequation
\label{dfg-1a22b593dbe7}
\begin{array}{@{}l}TKE\;=\;\frac{(0.1189\;\pm\;0.0011)Z^{2}}{A^{1/3}}+7.3\;\pm\;1.5\;MeV\end{array}
\end{eqnarray}
\global\let\theequation\saveeqnno
\addtocounter{equation}{-1}\ignorespaces 

derived by Viola, et al. \cite{1} where the second term represents the conversion of energy stored in deformation to kinetic energy.

\subsection{Previous studies of of $^{237} Np (n,f)$}
While the dependency of TKE on thermal neutron-induced fission has been extensively studied in many actinide systems, the same cannot be said with high incident neutron energies. There have been a limited number of studies on the $^{237}$ Np (n,f)\ reaction \cite{2,3} in which TKE release and mass distributions were reported. These studies are summarized in Table 1. Until this work, there have been no measurements made on the $^{237}$ Np (n,f)reaction for En > 5.5 MeV. As such the measurements presented in this paper substantially extend our knowledge of the TKE release for the $^{237} $Np (n,f) reaction.  

\begin{table}[!htbp]
\caption{{Previous work on the 237Np (n,f) reaction.} }
\label{table-wrap-93af179abb8747a0978484aea4b9e73d}
\begin{ruledtabular}
\ignorespaces 
\centering 
\begin{tabular}{ll}

\textbf{E}\ensuremath{_{\textbf{n}}} \textbf{ range (MeV)} &
  \\
 0.3 - 5.5 &
  \unskip~\cite{3}\\
 0.8 and 5.5 &
  \unskip~\cite{2}\\

\end{tabular}\par 
\end{ruledtabular}
\end{table}
\section{EXPERIMENTAL}
This experiment was carried out using the 15R beamline at Los Alamos National Laboratory-Los Alamos Neutron Science Center at the Weapons Neutron Research (LANL-LANSCE-WNR) facility over a 7 day period in December 2019. The LANSCE-WNR provides a high fluence of “white spectrum” neutrons generated by spallation of a thick tungsten target by 800 MeV protons, producing ~ $10^5 – 10^6$ n/s up to En = 100 MeV \cite{4, 5}. The proton beam consists of  625 $\mu$s macropulses. Each macropulse is composed of 250 ps-wide micropulses that are spaced 1.8 $\mu$s apart. The macropulses had a repetition rate of 100 Hz. The pulsed nature of the proton beam enables the measurement of the time-of-flight (energy) of the incident neutrons. \newline
\indent The general arrangement of the experiment is shown in Figure 1. The neutron flight path i.e. the distance from the spallation source to the $^{237}$Np target, was measured to be 13.85 m.  The neutron beam was collimated to a 1 cm diameter. An evacuated, thin-walled aluminum scattering chamber held the $^{237}$Np target and the Si PIN diode fission detectors. At the end of the 7 day irradiation period, a total of 8,618 coincident fission events were detected. 

\begin{figure*}
 \includegraphics[angle=-90,width=.75 \textwidth, trim=6cm 1.5cm 3cm 1.3cm, clip=true]{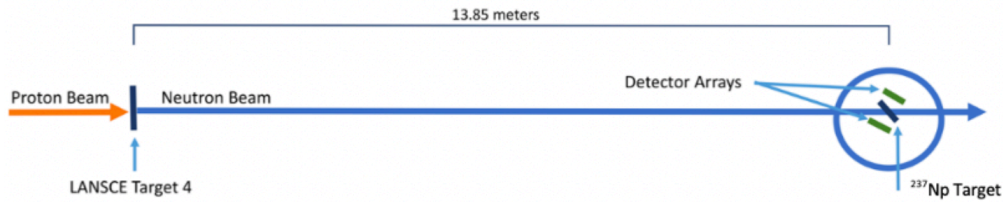}
 \centering
\vspace{-	2 cm}
\caption{Schematic diagram of the detection set-up at the LANSCE facility}
\label{fig:1}       
\end{figure*}

\subsection{Fission Detectors}
The scattering chamber was aligned with the neutron beam using a combination of neutron-exposed photo plates and an alignment laser. Inside the scattering chamber, two arrays of nine Si PIN photodiodes (Hamamatsu S3590-09), arranged in a 3x3 array, were used to detect the fission products. The arrays were arranged on opposite sides of the beam at angles of $65^o$ and $115^o$ from the incident beam. Each individual PIN diode has an area of 1 $cm^2$, and the detectors are placed 3.5 cm from the actinide target. Coincident fission fragments in both the left and right array were required to process a signal. The pulse height defect of the detectors for fission fragments was determined by employing the Schmitt method \cite{6} with a ${^252}$Cf source. The fragment energy loss in the actinide deposition and the carbon foil target backing was determined by Northcliff- Schilling energy loss tables on an event by event basis \cite{7}.\newline
\indent The time of flight (TOF) of each interacting neutron was measured using a timing pulse from a Si PIN diode and the accelerator RF signal. The neutron energies were calculated based on the relativistic relationship between the distance traveled by the neutrons, the position of the photo fission peak in the fission time of flight spectrum, and the observed time difference between the neutron timing signal and the accelerator RF signal \cite{8}. The position of the photofission peak was measured to have an uncertainty of 0.24 ns, with a relative uncertainty of 0.52\%. The 1.8 $\mu$s proton pulse rate allows for a small fraction of “wrap around” neutrons to reach the scattering chamber among the neutrons of the previous pulse. Accordingly, these neutrons would either be of very high energy, or very low energy. For the 13.85 m flight-path used in this experiment, the low-energy “wrap around” neutrons would have En $<$ 0.4 MeV. The near-negligible cross section for the $^{237}$Np(n, f) reaction at this energy means that these low-energy “wrap around” neutrons were ignored. Similarly, the high-energy “wrap around” neutrons would be of sufficient energy as to appear with the slowest neutrons from the previous beam burst. Wrap around events of this nature set the lower limit for the neutron energy resolution. Since no events below the fission threshold of En $\leq$ 0.7 MeV \cite{9} were detected, the high energy “wrap around” neutrons were also ignored. The neutron energies were binned logarithmically to give bins of equal associated uncertainty for the neutron energy.
\subsection{The 2E Method}
The 2E method is used to measure the fission product masses and their TKE by detecting the fission products and determining their energies after neutron emission. (We use the convention that the term “fission fragments” refers to pre-neutron emission while the term “fission product refers to post neutron emission.) Conservation of momentum and mass for the coincident fission products gives: 
let\saveeqnno\theequation
\let\savefrac\frac
\def\dispfrac{\displaystyle\savefrac}
\begin{eqnarray}
\let\frac\dispfrac
\gdef\theequation{3}
\let\theHequation\theequation
\label{dfg-afdea13e84b0}
\begin{array}{@{}l}M_H^\ast V_H^\ast\;=\;M_L^\ast V_L^\ast\end{array}
\end{eqnarray}
\global\let\theequation\saveeqnno
\addtocounter{equation}{-1}\ignorespaces 

\let\saveeqnno\theequation
\let\savefrac\frac
\def\dispfrac{\displaystyle\savefrac}
\begin{eqnarray}
\let\frac\dispfrac
\gdef\theequation{4}
\let\theHequation\theequation
\label{dfg-ba6bb0ee3049}
\begin{array}{@{}l}M_L^\ast\;+\;M_H^\ast\;=\;M_{CN}\end{array}
\end{eqnarray}
\global\let\theequation\saveeqnno
\addtocounter{equation}{-1}\ignorespaces 

\let\saveeqnno\theequation
\let\savefrac\frac
\def\dispfrac{\displaystyle\savefrac}
\begin{eqnarray}
\let\frac\dispfrac
\gdef\theequation{5}
\let\theHequation\theequation
\label{dfg-e48bfbe4ac93}
\begin{array}{@{}l}M_H^\ast E_H^\ast\;=\;M_L^\ast E_L^\ast\end{array}
\end{eqnarray}
\global\let\theequation\saveeqnno
\addtocounter{equation}{-1}\ignorespaces 

\let\saveeqnno\theequation
\let\savefrac\frac
\def\dispfrac{\displaystyle\savefrac}
\begin{eqnarray}
\let\frac\dispfrac
\gdef\theequation{6}
\let\theHequation\theequation
\label{dfg-8bb973e37bcf}
\begin{array}{@{}l}M_L^\ast\;=\;M_{CN}\;\frac{E_H^\ast}{E_H^\ast+E_L^\ast}\end{array}
\end{eqnarray}
\global\let\theequation\saveeqnno
\addtocounter{equation}{-1}\ignorespaces 
where H represents the heavy fragment, L represents the light fragment, CN represents the compound nucleus, and ‘indicates a quantity that pertains to pre-neutron emission. Our detection apparatus, due to the sub-femtosecond timescale of the compound nucleus’ descent to scission, measures post neutron evaporation fission products \cite{10}. Determination of mass yields requires pre-neutron evaporation masses. As such, we must have a sensible approximation for prompt neutron multiplicity as a function of excitation energy and post neutron fragment mass. We therefore make the following corrections for neutron emission.

\let\saveeqnno\theequation
\let\savefrac\frac
\def\dispfrac{\displaystyle\savefrac}
\begin{eqnarray}
\let\frac\dispfrac
\gdef\theequation{7}
\let\theHequation\theequation
\label{disp-formula-group-60bdec3559a94c37b6252425a37d1dad}
\begin{array}{@{}l}E_{L,H}^\ast\;=\;E_{L,H}\;\left(1+\frac{\nu_{post}(m_{L,H}^\ast)}{m_{L,H}}\right)\;\end{array}
\end{eqnarray}
\global\let\theequation\saveeqnno
\addtocounter{equation}{-1}\ignorespaces 

\let\saveeqnno\theequation
\let\savefrac\frac
\def\dispfrac{\displaystyle\savefrac}
\begin{eqnarray}
\let\frac\dispfrac
\gdef\theequation{8}
\let\theHequation\theequation
\label{disp-formula-group-065cf2a2eb3a4b5abedc5c2afbd3a7bb}
\begin{array}{@{}l}m_{L,H\;}=\;m_{L,H}^\ast\;-\;\nu_{post\;L,H}\;\left(E_{n,}m_{L,H}^\ast\right)\end{array}
\end{eqnarray}
\global\let\theequation\saveeqnno
\addtocounter{equation}{-1}\ignorespaces 

where ${v_{post} (A)}$ is the post fission prompt neutron multiplicity calculated from the General Description of Fission Observables (GEF) model \cite{9, 11}. In the laboratory system post neutron emission fission product energies $(E_{L,H})$ are calculated by correcting the measured fission product energies for detector pulse height defects and fission product energy losses in the target material and backing \cite{8}. To deduce the post neutron emission mass distributions an iterative correction procedure \cite{12} is applied. This iterative procedure assumes zero charged particle emission and ceases when the differences in fragment mass between iterations is less than 0.1 amu.

\subsection{Corrections for PHD and Fragment energy loss}
We measure, on an event by event basis, the pulse heights of the coincident fission fragments. To accurately transform these pulse heights into energies, we must account for the pulse height defect (PHD). A PHD occurs when a fission fragment strikes the PIN diode with such high energy that recombination occurs in the tracks of the incident particle along with any losses of energy in the detector dead layer. To make corrections for this effect we apply the Schmitt method \cite{6}. The Schmitt method exploits the known fission spectra of $^{252}$Cf to determine a set of constants the describe the PHD for a given detector. For Si surface barrier detectors, the magnitude of the PHD generally ranges from 3-13 MeV \cite{13,14,15} The energy loss of the fission products in the target deposit and the carbon foil were calculated using the Northcliffe-Schilling tables \cite{7}.

\begin{table*}[!htbp]
\caption{{TKE and $_{post}$TKE in the $^{237}$Np (n,f) reaction as a function of incident neutron energy. The neutron bin limits are given in the first column while the second column is the geometric mean of the neutron energies. The last column is the number of events N in the bin.} }
\label{table-wrap-67ef239583c14a9fa1a524d36be95d2f}
\begin{ruledtabular}
\ignorespaces 
\centering 
\begin{tabular}{llllll}

\textbf{E}\ensuremath{_{\textbf{n}}} \textbf{ range (MeV)} &
  \textbf{E}\ensuremath{_{\textbf{n}}} \textbf{(MeV)} &
  \textbf{\ensuremath{_{post}}} \textbf{TKE(MeV)} &
  \textbf{\ensuremath{_{pre}}}\textbf{TKE} \textbf{(MeV)} &
  \textbf{s}\textbf{\ensuremath{^{2}}}\textbf{\ensuremath{_{post}}}\textbf{(MeV}\textbf{\ensuremath{^{2}}} \textbf{)} &
  \textbf{Events}\\
 2.60-3.31 &
   2.96 &
   170.8 \ensuremath{\pm} 0.4 &
   173.1 \ensuremath{\pm} 0.4 &
   92.78 \ensuremath{\pm} 0.33 &
   887\\
 3.32-4.38 &
   3.83 &
   171.6 \ensuremath{\pm} 0.4 &
   173.8 \ensuremath{\pm} 0.4 &
   91.36 \ensuremath{\pm} 0.33 &
   868\\
 4.39-6.09 &
   5.19 &
   171.0 \ensuremath{\pm} 0.4 &
   173.6 \ensuremath{\pm} 0.4 &
   89.40 \ensuremath{\pm} 0.33 &
   865\\
 6.10-8.06 &
   7.06 &
   171.2 \ensuremath{\pm} 0.4 &
   173.3 \ensuremath{\pm} 0.4 &
   109.83 \ensuremath{\pm} 0.35 &
   861\\
 8.07-11.66 &
   9.7 &
   170.2 \ensuremath{\pm} 0.4 &
   172.2 \ensuremath{\pm} 0.4 &
   107.95 \ensuremath{\pm} 0.34 &
   867\\
 11.67-18.56 &
   14.8 &
   168.0 \ensuremath{\pm} 0.4 &
   170.9 \ensuremath{\pm} 0.4 &
   109.62 \ensuremath{\pm} 0.42 &
   861\\
 18.57-29.66 &
   23.9 &
   167.1 \ensuremath{\pm} 0.4 &
   169.5 \ensuremath{\pm} 0.4 &
   107.33 \ensuremath{\pm} 0.35 &
   863\\
 29.67-45.72 &
   37.1 &
   164.9 \ensuremath{\pm} 0.4 &
   167.5 \ensuremath{\pm} 0.4 &
   118.85 \ensuremath{\pm} 0.39 &
   862\\
 45.73-68.78 &
   56.5 &
   166.0 \ensuremath{\pm} 0.4 &
   168.6 \ensuremath{\pm} 0.4 &
   113.85 \ensuremath{\pm} 0.36 &
   870\\
 68.79-100 &
   83.8 &
   164.8 \ensuremath{\pm} 0.5 &
   168.0 \ensuremath{\pm} 0.4 &
   126.34 \ensuremath{\pm} 0.44 &
   814\\

\end{tabular}\par 
\end{ruledtabular}
\end{table*}

\begin{figure}
 \includegraphics[width=.5\textwidth]{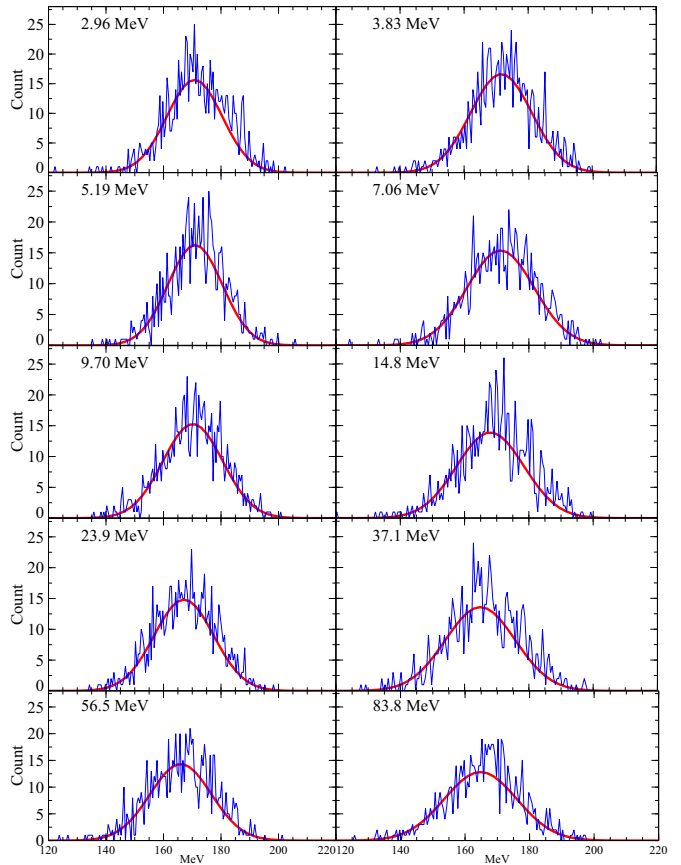}
\caption{Plots for the post TKE distributions sorted into the energy bins described in Table 2. The plotted curves (line) represent the results of fitting the distributions with Gaussian functions. (Color online).}
\label{fig:2}       
\end{figure}

\section{Results}
In Table 2, we report the pre- and post-neutron TKE and its variance as a function of incident neutron energy for the $^{237}$Np (n,f) reaction. This data set significantly increases the range of energies studied compared to previous measurements that only explored this reaction to a maximum value of 5.5 MeV in neutron energy \cite{2,3}. In  Figure 2, we present the post neutron evaporation TKE distributions sorted by neutron energy bin. The data were binned as to have a similar number of events per neutron energy bin. The thin red lines in each plot represent the results of fitting the data with Gaussian distributions.

\subsection{TKE vs $E_n$}
 In Figure 3, we show the $_{post}$TKE response up to En = 6 MeV from this work and that documented in the literature \cite{2, 3}. In doing so, we restrict our attention to systems where only first chance fission is expected. The experiment found the TKE released in the low En energy range to be ~1.7 and 0.9 MeV higher than that reported by Hambsch \cite{3} and Naqvi \cite{2} respectively. It is important to note that Hambsch normalized their data to a $^{235}$U pre-neutron TKE of 179.5 MeV, a value ~0.5 MeV lower than the compilation of Madland \cite{16}. Similarly, Naqvi compared their $^{237}$Np TKE data to a $^{235}$U post-neutron TKE value of 167.45 MeV. This value is 1.7 MeV lower than what is reported by Madland. As a result, both the Naqvi and Hambsch data likely underestimate the TKE release in $^{237}$Np. \newline
 \indent The $_{post}$TKE release for the fast neutron-induced fission of $^{237}$Np over the entire measured energy range is given in Figure 4. In Figure 4, we show our data, as well as comparisons to other $^{237}$Np (n,f) reaction data and the GEF model. The overall trend of decreasing TKE with increasing excitation is consistent with other fast neutron TKE studies of actinides \cite{17,18,19,20,21,22,23,24,25,26}. While the TKE release for the lowest energy neutrons is approximately linear, a first order log$_{10}(En)$ polynomial fit is needed to describe the fast neutron energy region (En$ >$ 1 MeV).  The fit to our data (Figure 4) is given by the equation TKE(MeV) = (174.38$\pm$  0.72) – (5.11 $\pm$ 0..5821) log$_{10}$ En. \newline
 \indent An interesting feature apparent in Figure 4 is a “bump” in TKE in the En   = 46-69 MeV energy bin. This peculiar bump has been observed in other actinides within this neutron energy range, particularly with $^{232}$Th, $^{235}$U and $^{239}$Pu \cite{17, 18, 25}. This anomaly is not predicted in models. The observation of such a “bump” across several actinides suggests either a systematic flaw in our measurement or a gap in our understanding of nuclear physics in this energy range.
 
 \subsection{$_{post}$TKE comparison to the GEF model}
When comparing the GEF model postTKE values to those measured in this work, GEF systematically overestimates postTKE by ~1.2$\%$ for all data points in the region En $\leq$ 50 MeV. This is equivalent to approximately 2 MeV per energy bin (Figure 4). This overestimation is likely the result of the GEF model misattributing a small fraction
of the energy of the fissioning $^{237}$Np nucleus to the kinetic energy of the fission fragments instead of particle evaporation. A study of the high energy fission of $^{235}$U(n,f) found that the  GEF framework routinely underestimated the total neutron evaporation by 5-10$\%$  for En = 0- 50 MeV \cite{27}. A similar underestimation is seen for the $^{239}$Pu (n,f) reaction, where the GEF prompt neutron multiplicity underestimates the known values for A $>$ 147 amu, and/or En $>$ 12 MeV \cite{28,29,30,31,32}. \newline
\indent The GEF model underestimates the $_{post}$TKE for En $>$ 50 MeV by ~ 0.7$\%$ or 1.1 MeV. This transition from the GEF model overestimation to sudden underestimation is due to the “bump” in the experimental data and further highlights the importance of being able to understand this anomaly. 

\begin{figure}
 \includegraphics[width=.5\textwidth]{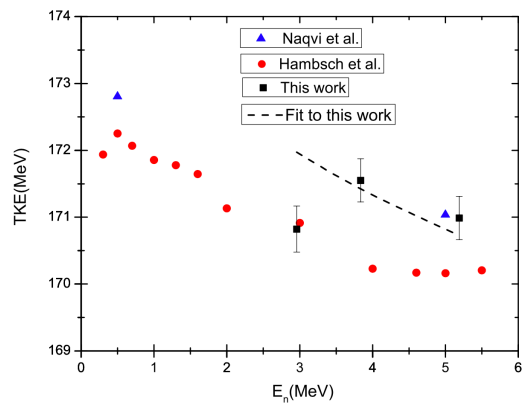}
\caption{Comparison of the average postTKE measured for the 237Np (n,f) reaction as a function of incident neutron energy measured in this work to previous studies and a fit to our data (dashed line)(Color-online)}
\label{fig:3}       
\end{figure}

\begin{figure}
 \includegraphics[angle=-90,width=.5\textwidth]{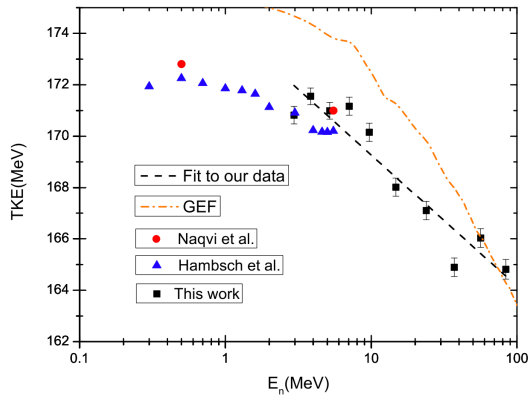}
 \vspace{-1cm}
\caption{Post neutron TKE released for the 237Np (n,f) reaction from En =  2.6 -100 MeV. TKE error bars are statistical (Color-online)}
\label{fig:4}       
\end{figure}

\subsection{TKE distribution Variances ($\sigma$)}
In Figure 2, the TKE data are fit with Gaussian distributions. The variances of the Gaussian fits, given in Table 2, are sensitive markers of multi-chance fission. At energies where multi-chance fission is possible, a new fission pathway opens, and the presence of this new channel is reflected in the broadening of the Gaussian distribution. This phenomenon is well documented, particularly in $^{235}$U (n,f) studies \cite{17,33}. In Figure 5, we limit our attention to measured variances below 20 MeV to highlight the dependence of the measured variances of the TKE distributions on the number of fission channels present. As expected, we see a significant increase in the measured variance at the onset of second chance fission, ~7 MeV. 

\begin{figure}
 \includegraphics[width=.5\textwidth, trim=2cm 1.5cm 1cm 1.3cm, clip=true]{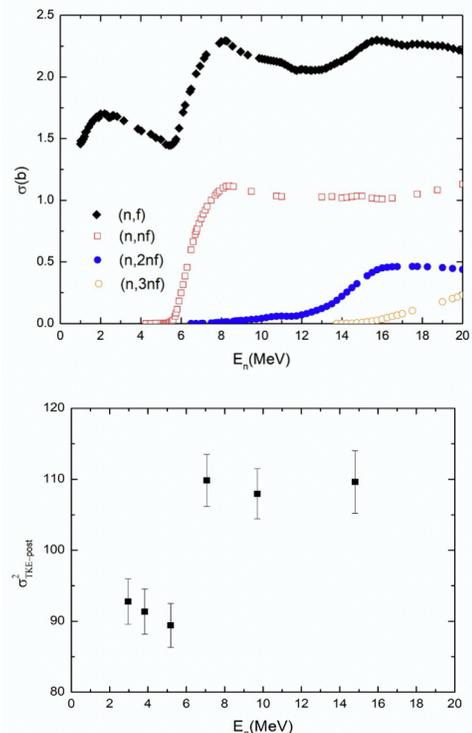}
\vspace{-2cm}
\caption{Top: cross section for $^{237}$Np (n,f) from ENDF/B-VIII, ENDF/B-VI. Bottom: measured $TKE  \sigma^2$ vs En for this study. (Color online)}
\label{fig:5}       
\end{figure}

\subsection{Mass Yield Distributions}
In actinide fission, the fission fragment mass distribution is bimodal at low excitation energies. As E$_n$ is increased the valley between the asymmetric fission peaks begins to fill. In Figure 6 we plot the mass distributions measured in this work, normalized to 200$\%$,  and the predictions of the GEF model. The resolution of our measurement is $\pm$ 5 amu. It is evident in  Figure 6 that as End increases the two asymmetric peaks merge reflecting the emergence of symmetric fission. The rate of symmetric ingrowth, and its relative proportion to the total number of events, is in excellent agreement with that predicted by GEF \cite{9}. \newline
\indent It is well documented in the literature \cite{34} that the constancy of the heavy fragment mass peak is due to the influence of the spherical N=82 and deformed N=88 neutron shells and Z = 50 proton shell. In this system the mass of the heavy fragment is anchored at ~ 139.4 amu. This is in agreement with \cite{3}.

\begin{figure}
 \includegraphics[angle = 90, width=.55\textwidth]{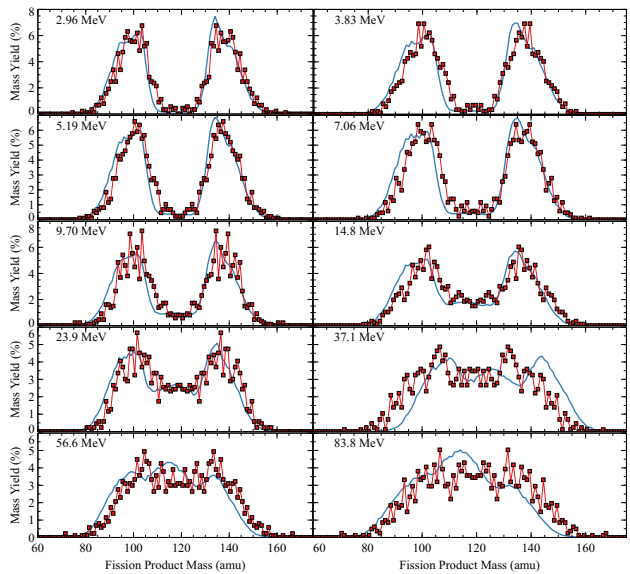}
\caption{Comparison of the mass distributions measured in this work (red squares), normalized to 200$\%$,  and the predictions of the GEF$_{post}$  model (blue line)\cite{9}  (Color-online)}
\label{fig:6}       
\end{figure}
 
 \subsection{Fission Channel Symmetry and TKE}
 It has been demonstrated in many actinide systems that the general trend of decreasing TKE with increasing excitation is a consequence of the increasing probability of symmetric fission as En increases. For low energy neutron-induced fission, asymmetric fission dominates due to nuclear shell effects. As En increases, symmetric fission begins to emerge, and eventually, symmetric fission dominates. Figure 7 shows the fragment mass and TKE distributions for $^{237}$Np at various energies. At the lowest neutron energies, the fissioning system is almost purely asymmetric. As En increases, more fission fragments with A = 120 amu are detected indicating a symmetric fission event. 
 
 \begin{figure}
\includegraphics[angle=-90,width=.5 \textwidth]{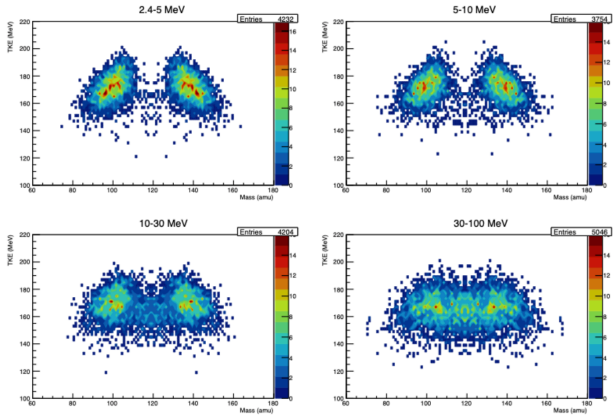}
\caption{$^{237}$Np (n,f) product  mass TKE distributions at various En.  (Color online)}
\label{fig:7}       
\end{figure}

The symmetry of the fissioning system was determined via the calculated masses of the fission products. For an event to fall into the asymmetric bin, the following condition had to be meant for one of the fission fragments, 
\let\saveeqnno\theequation
\let\savefrac\frac
\def\dispfrac{\displaystyle\savefrac}
\begin{eqnarray}
\let\frac\dispfrac
\gdef\theequation{9}
\let\theHequation\theequation
\label{disp-formula-group-67de663e2f3b460bb818280cfe82ae45}
\begin{array}{@{}l}A_{range}\;=\;\left(\frac{A_{CN}\;-\;v_{tot,En}}2\;+\;20\right)\pm\;5\;amu\end{array}
\end{eqnarray}
\global\let\theequation\saveeqnno
\addtocounter{equation}{-1}\ignorespaces 
where $_{vtot,En}$ is the total neutron multiplicity as calculated by GEF \cite{9} at the corresponding En. The second term (+ 20) is the calculated difference between the A$_H$  peak in asymmetric fission and the single peak corresponding to symmetric fission. If both fission products fall within a finite mass range ($S_{range}$) the event passed into the symmetric bin, where $S_{range}$  is defined as,

\ensuremath{_{}}is defined as,
\let\saveeqnno\theequation
\let\savefrac\frac
\def\dispfrac{\displaystyle\savefrac}
\begin{eqnarray}
\let\frac\dispfrac
\gdef\theequation{10}
\let\theHequation\theequation
\label{disp-formula-group-80ec323a08ab4b359c054f9a64b27603}
\begin{array}{@{}l}\begin{array}{l}\begin{array}{l}S_{range}\;=\;\frac{A_{CN}\;-\;v_{tot,En}}2\;\pm\;5\;amu\\\\\end{array}\\\end{array}\end{array}
\end{eqnarray}
\global\let\theequation\saveeqnno
\addtocounter{equation}{-1}\ignorespaces 

\begin{figure}
 \includegraphics[width=.5 \textwidth, trim=5cm 1.5cm 3cm 4.3cm, clip=true]{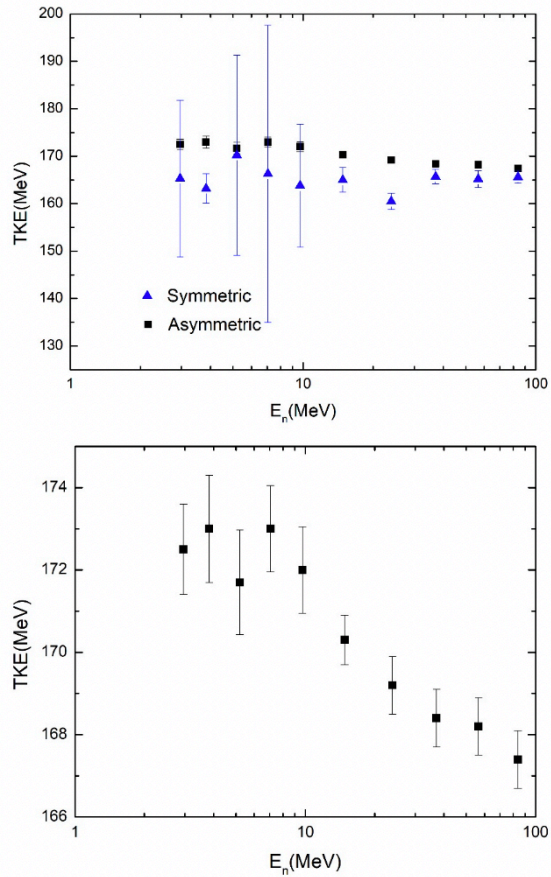}
 \vspace{-3cm}
\caption{(a)  TKE vs. Em with fission events separated into symmetric and asymmetric events. (b)Dependence of asymmetric TKE events on neutron energy.(Color online)}
\label{fig:8}       
\end{figure}

In Figure 8, we show the TKE$_{sym}$ and TKE$_{asym}$ events and plot them against En.  In Figure 8 (inset) we  see that TKE$_{asym}$ drops off rather sharply above En = 10 MeV. Across the entire En range measured, TKE$_{asym}$ decreased from 172.5 MeV to 165.6 MeV. However , the TKEs release remained virtually constant. This suggests that the overall trend of decreasing TKE is not exclusively a consequence of the onset of symmetric fission but is instead also impacted by a decreasing TKEasym as well. Hambsch theorized that this decreasing asymmetric TKE trend observed in the$^{237}$Np(n,f) reaction could involve a more drastic deformation of fission fragments, particularly for 130 $<$ A $<$ 145 amu range, as En increases \cite{3}.

\section{CONCLUSION}
Previous TKE studies of $^{237}$Np (n,f) are limited and were restricted to  incident neutron energies less than 6 MeV. The goal of this work was to build upon the existing TKE data available for $^{237}$Np by extending measurements into the fast neutron energy range, up to 100 MeV. \newline
\indent A vapor deposited $^{237}$NpF$_4$ target was irradiated at LANSCE-WNR to measure the TKE dependence for the $^{237}$Np (n,f) reaction over  En 2.6 -100 MeV. The TKE values  reported in this work were compared to that of \cite{2,3} and the prediction of the GEF model \cite{9}. We also report fission fragment mass distributions and gained additional detail about the fissioning system by separating the fission path by symmetry regimes. We attribute the observed trend of decreasing TKE to the emergence of symmetric fission, as well as the rapid drop off of TKE$_{asym}$ as excitation increases. The peculiar “bump” in TKE at En = 45-70 MeV, while consistent with other actinide studies within this framework, is not yet understood and needs further exploration.

\end{document}